\documentclass{article}

    \usepackage[final, nonatbib]{neurips_2021}

\usepackage[utf8]{inputenc} 
\usepackage[T1]{fontenc}    
\usepackage{hyperref}       
\usepackage{url}            
\usepackage{booktabs}       
\usepackage{amsfonts}       
\usepackage{nicefrac}       
\usepackage{microtype}      
\usepackage{graphicx}
\usepackage{wrapfig}
\usepackage{caption}
\usepackage{makecell}
\usepackage{amsmath}

\title{Deciphering antibody affinity maturation with language models and weakly supervised learning}

\author{%
  Jeffrey A. Ruffolo \\
  Johns Hopkins University \\
  Baltimore, MD 21218 \\
  \texttt{jruffolo@jhu.edu}
   \And
   Jeffrey J. Gray \\
   Johns Hopkins University \\
   Baltimore, MD 21218 \\
   \texttt{jgray@jhu.edu}
   \And
   Jeremias Sulam \\
   Johns Hopkins University \\
   Baltimore, MD \\
   \texttt{jsulam@jhu.edu}
}

\begin{document}

\maketitle

\begin{abstract}
In response to pathogens, the adaptive immune system generates specific antibodies that bind and neutralize foreign antigens.  Understanding the composition of an individual's immune repertoire can provide insights into this process and reveal potential therapeutic antibodies.  In this work, we explore the application of antibody-specific language models to aid understanding of immune repertoires.  We introduce AntiBERTy, a language model trained on 558M natural antibody sequences.  We find that within repertoires, our model clusters antibodies into trajectories resembling affinity maturation.  Importantly, we show that models trained to predict highly redundant sequences under a multiple instance learning framework identify key binding residues in the process.  With further development, the methods presented here will provide new insights into antigen binding from repertoire sequences alone.
\end{abstract}

\section{Introduction}

The adaptive immune system is capable of generating robust responses to foreign pathogens.  This robustness is provided in part by the immense diversity of antibodies that can be generated.  Diversity is initially introduced to antibody sequences through V(D)J gene recombination.  As the immune response progresses, antibodies capable of effective neutralization are developed through an antigen-driven process called affinity maturation.  In this process, B-cells producing antibodies with high antigen affinity are selectively expanded then mutated to give rise to successive generations of antibodies.

Immune repertoire samples provide a snapshot of an individual's antibody sequence population.  Typically, donors provide samples of B-cells from blood or lymph, and the antibodies produced by these B-cells are identified via next-generation sequencing \cite{georgiou2014promise}.  During an immune response, as many as half of the antibodies within the repertoire may exhibit antigen affinity \cite{neumeier2021phenotypic}.  Among these binding antibodies, the most frequently occurring sequences tend to be effective binders \cite{reddy2010monoclonal}.  However, beyond the small subset of highly expanded sequences, redundancy is a poor indicator of binding capability \cite{neumeier2021phenotypic}.

Models adopted from natural language processing and trained on massive sets of protein sequences have been shown repeatedly to learn rich representations of protein sequences \cite{elnaggar2020prottrans, rives2021biological}.  Such models have been used for mutational variant prediction \cite{meier2021language}, to generate embeddings for structure prediction \cite{chowdhury2021single}, and even to study evolution within protein families \cite{hie2021evolutionary}.  However, natural proteins evolve under many selective pressures, while antibodies are selected for binding to a particular antigen.  As such, models trained on all proteins may be poorly suited for capturing specific features of antibody sequence evolution.  In this work, we explore the use of an antibody-specific language model for understanding affinity maturation within immune repertoires.

\section{Methods}

\subsection{Antibody encoder model}

To learn antibody-specific representations, we trained a transformer encoder model \cite{devlin2018bert} on 558M non-redundant sequences from the Observed Antibody Space \cite{kovaltsuk2018observed} (Appendix~\ref{antibody_dataset}).  The model, which we call AntiBERTy, is based on the BERT architecture \cite{devlin2018bert} implementation from Huggingface \cite{wolf2019huggingface}.  We trained using the masked language modeling (MLM) objective, as originally proposed by Devlin et al. \cite{devlin2018bert}.  Specific model parameters and training loss plots are provided provided in Appendix~\ref{bert_params}.

\subsection{Evolutionary analysis of immune repertoires}

Inspired by recent work in modeling protein evolution with language models \cite{hie2021evolutionary}, we began by looking for global trends within individual repertoires.  We considered immune repertoire samples from four donors who developed neutralizing antibodies against HIV-1 \cite{zhou2013multidonor, zhou2015structural} (Appendix~\ref{repertoire_info}).  Specifically, these donors developed antibodies belonging to the VRC01 class, which bind to the HIV-1 spike glycoprotein120 (gp120) through a $V_H$-gene mediated paratope \cite{wu2010rational, zhou2015structural}.  For each repertoire, we created a \textit{k}-nearest-neighbor graph using embeddings from AntiBERTy as described by Hie et al \cite{hie2021evolutionary}.  We additionally plotted evo-velocity scores (Appendix~\ref{evolocity_scores}).

\begin{wrapfigure}{l}{0.4\textwidth}
    \captionsetup{width=\linewidth,skip=-40 pt,position=top}
    \centering
    \includegraphics[width=0.4\textwidth]{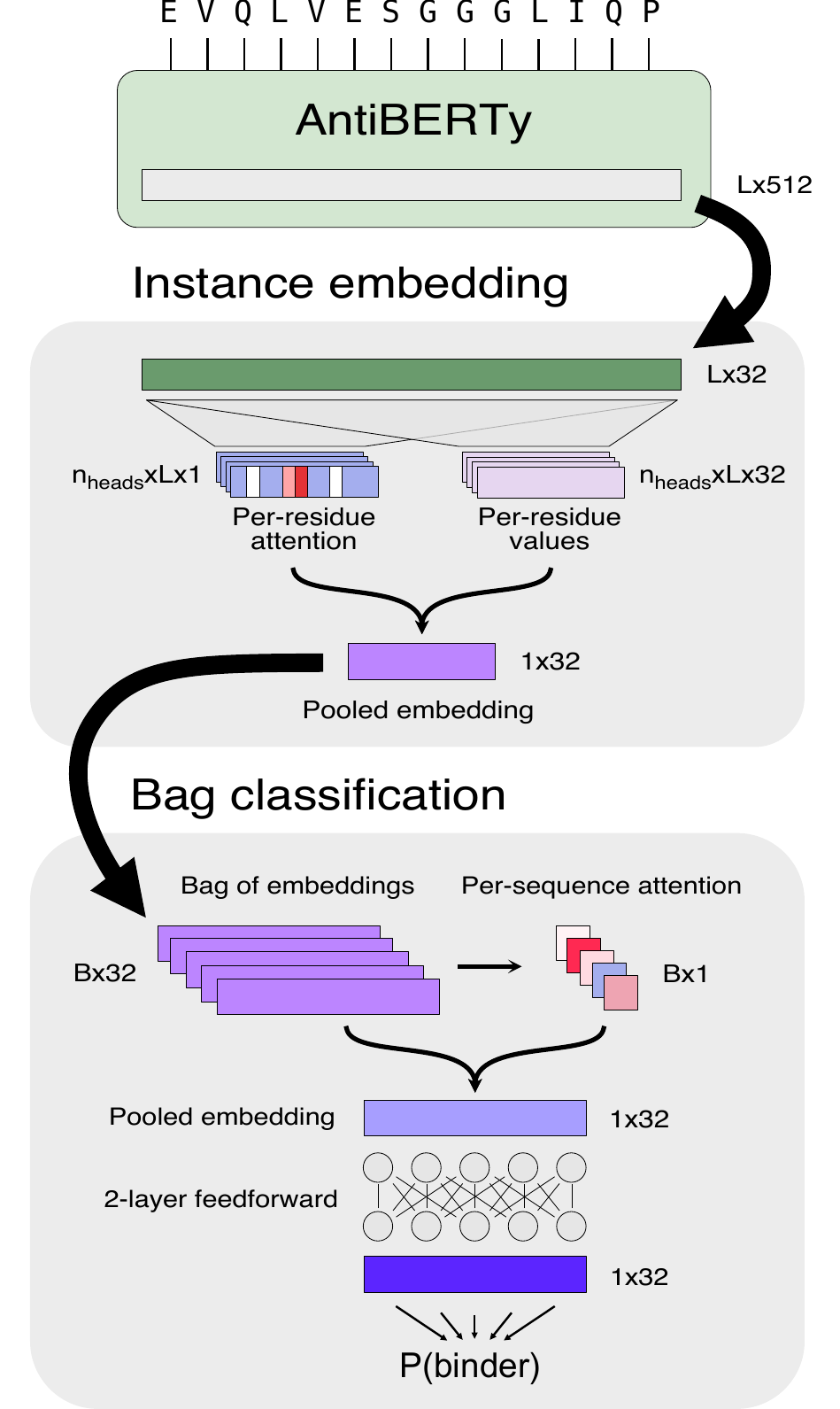}
    \caption{Diagram of MIL model for predicting whether a bag of sequences contains a highly redundant instance.}
    \label{fig:figure_1}
\end{wrapfigure}

\subsection{Multiple instance learning on sequence embeddings}

For a given sequence in a repertoire, we sought to identify the residues contributing to antigen binding (i.e., the paratope).  However, our objective is complicated by the absence of labels describing the binding capabilities of individual antibody sequences within the repertoire.  Instead, we rely on the connection between clonal expansion and antigen binding \cite{reddy2010monoclonal} to construct a noisy label – i.e., we assume that the most frequently observed antibodies are binders.  We then adopt a multiple instance learning (MIL) framework \cite{dietterich1997solving} to predict whether a set of sequences is likely to contain a highly redundant instance (and thus, binding residues).

\subsubsection{MIL dataset creation}

For each repertoire, we compute the 85th percentile of redundancy values and consider sequences with greater redundancy to be likely binders and those with lower redundancy to be unlikely binders.  Then, to generate training examples, we sample bags of 64 sequences (uniformly at random) from the more- or less-redundant partitions to create positive or negative examples, respectively.

\subsubsection{Model architecture and training}

We construct a MIL model consisting of two primary components (Fig.~\ref{fig:figure_1}): an instance embedding module and an MIL pooling classifier.  The inputs to the instance embedding module are generated by passing an amino acid sequence $S=(s_1,\dots,s_L)$ through AntiBERTy.  We extract the final hidden representation from AntiBERTy and reduce the dimensionality using a linear transformation, resulting in a variable-length embedding $H=(\mathbf{h}_1,\dots,\mathbf{h}_L)$ with $\mathbf{h}_i \in \mathbb{R}^{d_\text{emb}}$.  Then, we use a multi-head attention \cite{vaswani2017attention} pooling layer to reduce the variable-length embedding to a fixed size vector, described below for one attention head:

\begin{equation}
    {a}_i^\text{emb} = \frac{\exp ( \mathbf{w}_\text{emb}^T (\mathbf{Q} \mathbf{h}_i^T \bigodot \mathbf{K} \mathbf{h}_i^T ) )}{ \sum_j^L{\exp ( \mathbf{w}_\text{emb}^T (\mathbf{Q} \mathbf{h}_j^T \bigodot \mathbf{K} \mathbf{h}_j^T ) )}}
\end{equation}
in order to obtain the final fixed-size instance embedding $\mathbf{x} = \sum_i^L{a_i^\text{emb} \mathbf{V} \mathbf{h}_i^T}$.
In the above, $\mathbf{Q}, \mathbf{K}, \mathbf{V} \in \mathbb{R}^{d_\text{emb} \times d_\text{head}}$ and $\mathbf{w}_\text{emb} \in \mathbb{R}^{d_\text{head} \times 1}$ are learnable parameters, and $\mathbf{a}^\text{emb} \in \mathbb{R}^L$ is a per-residue attention score.  In practice, we use four attention heads and concatenate the fixed-size embeddings.  Each sequence in a bag is passed individually through the instance embedding module and the embeddings are collected to form the set $X=\{\mathbf{x}_1,\dots,\mathbf{x}_B\}$.  We adopt the gated attention mechanism from Ilse et al \cite{ilse2018attention} to learn a MIL pooling operation:

\begin{equation}
    a_i^\text{bag} = \frac{\exp ( \mathbf{w}_\text{bag}^T (\operatorname{sigm}(\mathbf{V} \mathbf{h}_i^T) \bigodot \operatorname{sigm}(\mathbf{U} \mathbf{h}_i^T) ) )}{ \sum_j^B{\exp ( \mathbf{w}_\text{bag}^T (\mathrm{sigm}(\mathbf{V} \mathbf{h}_j^T) \bigodot \mathrm{sigm}(\mathbf{U} \mathbf{h}_j^T) ) )}},
\end{equation}
finally obtaining
\begin{equation}
    \mathbf{z} = \sum_i^B{a_i^\text{bag} \mathbf{h}_i},
\end{equation}
where $\mathbf{V}, \mathbf{U} \in \mathbb{R}^{d_\text{attn}}$ and $\mathbf{w}_\text{bag} \in \mathbb{R}^{d_\text{attn} \times 1}$ are learnable parameters, $\mathbf{a}^\text{bag} \in \mathbb{R}^B$ is a per-instance attention score, and $\mathbf{z} \in \mathbb{R}^{d_\text{emb}}$ is a fixed-size bag embedding.  Finally, the bag classification is predicted with a two-layer feed-forward network followed by the logistic function.  

Specific model parameters are provided in Appendix~\ref{mil_params}.  We use cross-entropy loss to train the model for 20 epochs, where an epoch is defined as the total repertoire size divided by the bag size.  During training, we sample positive and negative bags with equal frequency.

\section{Results}

\subsection{Language model reveals trajectories within repertoire}

\begin{figure}[b]
    \captionsetup{width=\linewidth}
    \includegraphics[width=\textwidth]{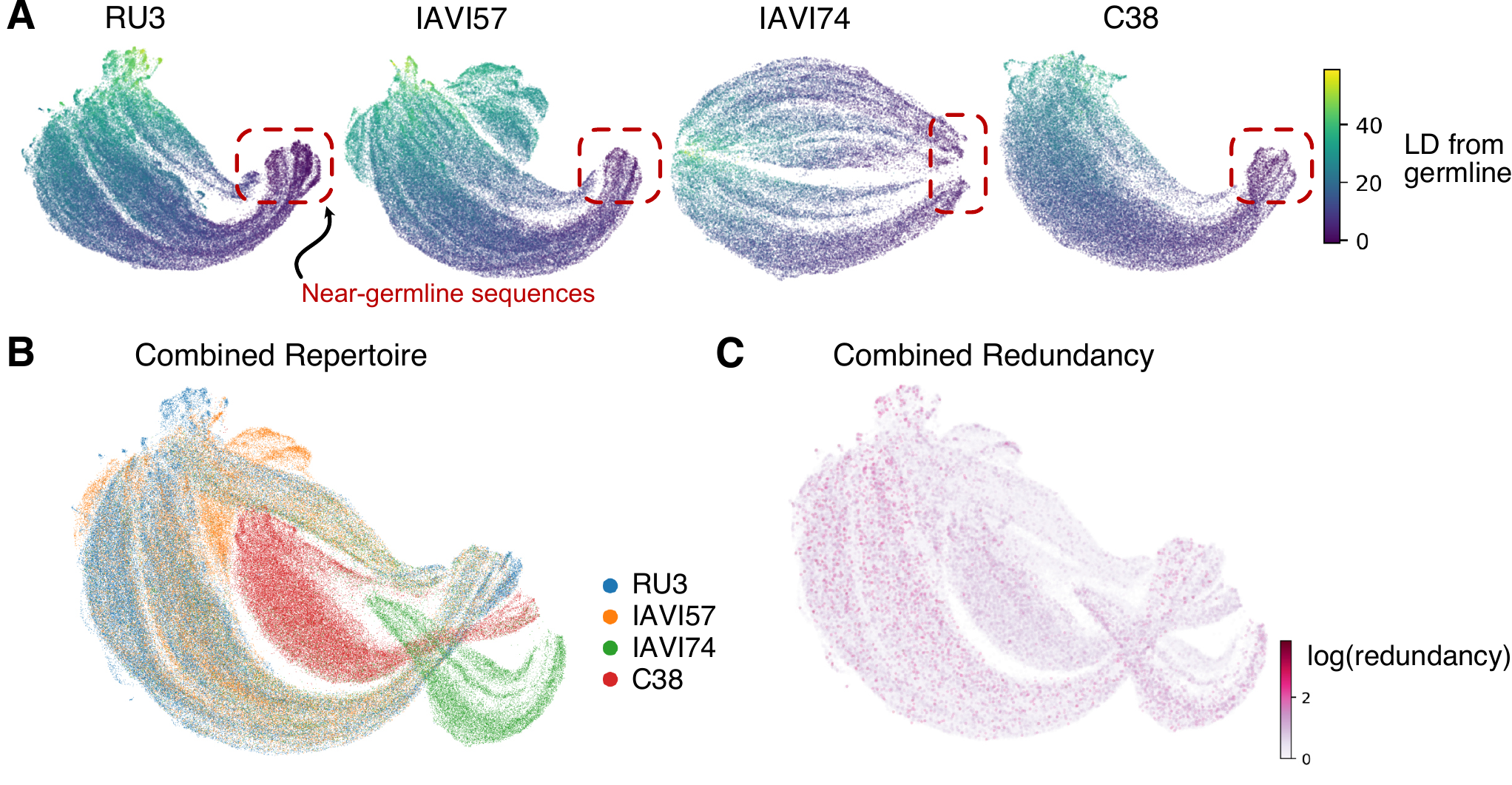}
    \caption{Evolutionary analysis of immune repertoires. (A) UMAP embedding of repertoire sequences annotated with Levenshtein distance (LD) from germline for four donors. (B) UMAP embedding of four-donor combined repertoire. (C) Combined repertoire annotated with sequence redundancy.}
    \centering
    \label{fig:figure_2}
\end{figure}

We applied the evolutionary analysis to repertoires samples from four donors who produced VRC01 class antibodies.  First, we visualized the KNN graph in two-dimensional UMAP \cite{mcinnes2018umap} embedding and annotated each sequence with the distance from germline (Fig.~\ref{fig:figure_2}A).  For each donor, we observe continuous trajectories between germline sequences and highly mutated derivatives, consistent with the process of affinity maturation.  Next, we combined the repertoires from all donors into a single set and repeated the analysis (Fig.~\ref{fig:figure_2}B).  The combined sequence graph displays significant overlap between the antibody sequences from donors RU3, IAVI57, and IAVI74, consistent with previous findings that VRC01 antibodies share ontogenies \cite{zhou2013multidonor}.  In constrast, the repertoire from donor C38 overlaps little with other donors, likely due to maturation from an alternative set of germline sequences.  Finally, we show the redundancy of each sequence within the combined repertoire (Fig.~\ref{fig:figure_2}C).  We observe a relatively uniform distribution of redundancy throughout the embedded space.  This uniformity is likely a reflection of iterative nature of affinity maturation, by which sequences are developed through many rounds of clonal expansion and diversification.

\begin{figure}[h]
    \captionsetup{width=\linewidth}
    \includegraphics[width=\textwidth]{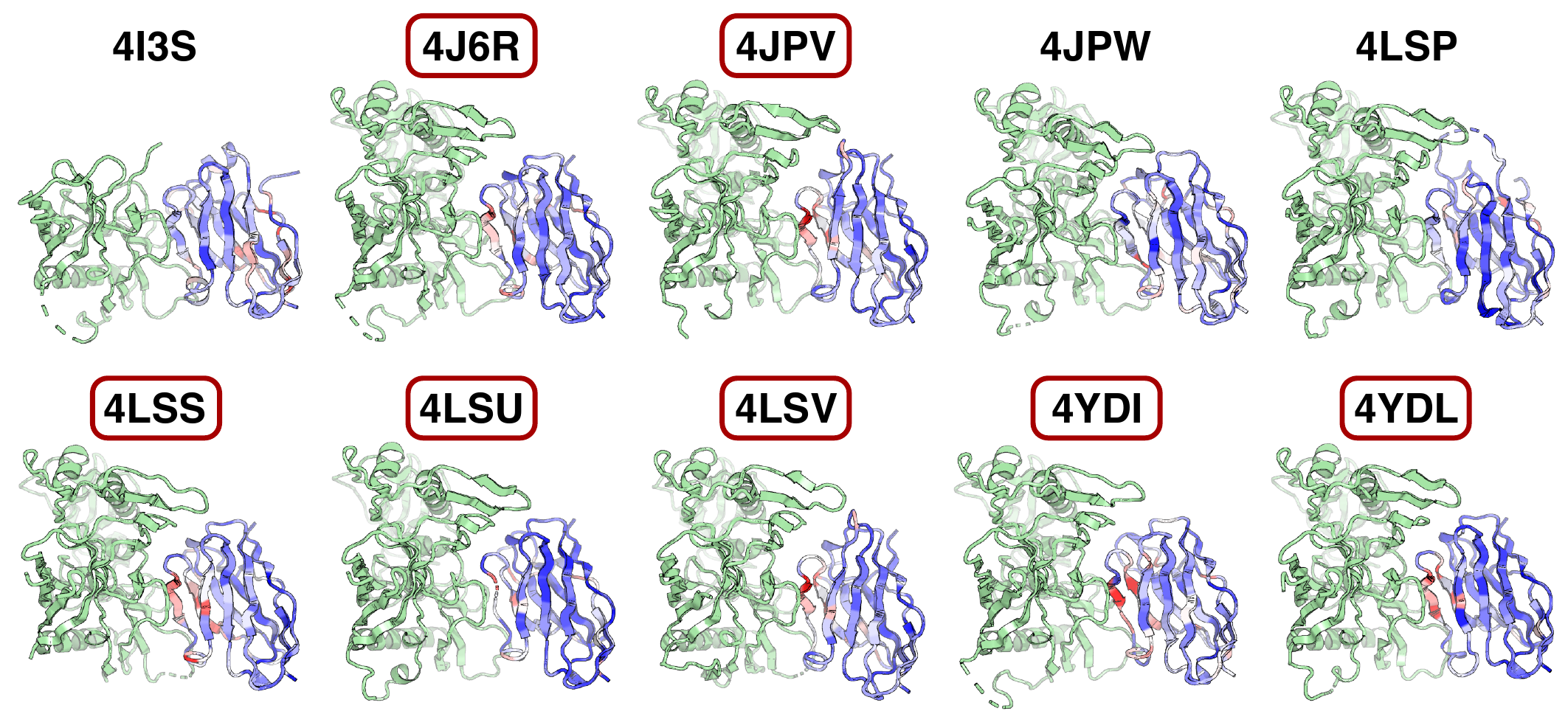}
    \caption{MIL model attention reveals paratope. Ten VRC01 antibodies annotated with residue-level attention from MIL model in complex with gp120 antigen (green). Attention values increase from blue to red. For seven antibodies (indicated with red boxes), attention localizes to paratope residues.}
    \centering
    \label{fig:figure_3}
\end{figure}

\subsection{MIL model identifies VRC01 paratope}

MIL models were trained for each individual repertoire, as well as the combined repertoire, as described above.  For each dataset, the model learned to effectively identify bags containing high-redundancy sequences (Appendix~\ref{mil_metrics}).  To verify that the model had learned properties of VRC01 antibodies, we collected a set of ten VRC01 antibodies crystallized in complex with the gp120 antigen \cite{joyce2013outer, klein2013somatic, zhou2015structural, zhou2013multidonor}.  We created single-instance bags for each sequence and confirmed  that the MIL model successfully produced positive predictions.  Next, we investigated whether the residues contributing to predictions were consistent with the binding mode of VRC01 antibodies.  For each complex, we annotated the antibody structure with the attention $\mathbf{a}^\text{emb}$ from each of four heads of the instance embedding module (Fig.~\ref{fig:figure_3}).  For seven of the ten sequences, the second attention head showed remarkable localization to the H2 loop and the following beta strand, consistent with the VRC01 paratope (Fig.~\ref{fig:figure_2}).  The remaining attention heads scattered attention throughout the structure (Appendix~\ref{mil_heads}), consistent with previous observations that substantial framework mutations are necessary for functional VRC01 antibodies \cite{klein2013somatic}.

\section{Conclusion}

In this work, we explored the use of language models to study the affinity maturation process.  Towards this goal, we developed AntiBERTy, an antibody-specific language model trained on a massive set of natural antibody sequences.  Next, we showed that immune repertoire sequences encoded with AntiBERTy cluster to resemble affinity maturation trajectories.  Closer inspection of these trajectories may provide new biological insights into the affinity maturation process.  Finally, we trained a MIL model to detect highly redundant sequences, and showed that the model’s attention localized to binding residues for an extensively studied class of antibodies.  In the future, similar methods may enable identification of paratope residues from immune repertoire sequences alone.

\begin{ack}
The authors thank Jacopo Teneggi, Zhenzhen Wang, Richard Shuai, Dr. Sai Pooja Mahajan, and Dr. Rahel Frick for helpful discussions and advice. This work was supported by National Institutes of Health grants R01-GM078221 and T32-GM008403 (J.A.R.), CISCO research grant CG\# 2686384, and AstraZeneca (J.A.R.). Computational resources were provided by the Maryland Advanced Research Computing Cluster (MARCC).
\end{ack}

\newpage

\bibliographystyle{plain}
\bibliography{citations}

\newpage
\appendix

\section{Appendix}

\subsection{Antibody sequence dataset}
\label{antibody_dataset}

The Observed Antibody Space is a database containing over 1B antibody variable domain sequneces from 80 immune repertoire sequencing studies.  We clustered this set at 95\% sequence identity with LinClust \cite{steinegger2018clustering} to extract a non-redundant set of 588M sequences.  Our dataset includes heavy and light chains from six species (human, mouse, rat, camel, rabbit, and rhesus). From the 588M sequences extracted from the OAS database, we hold out 5\% for future testing. Of the remaining 95\%, we train the model 558M sequences and use 1M for evaluation and hyperparameter tuning.

\subsection{AntiBERTy model}
\label{bert_params}

\begin{figure}[h]
    \captionsetup{width=\linewidth}
    \centering
    \includegraphics[width=0.75\textwidth]{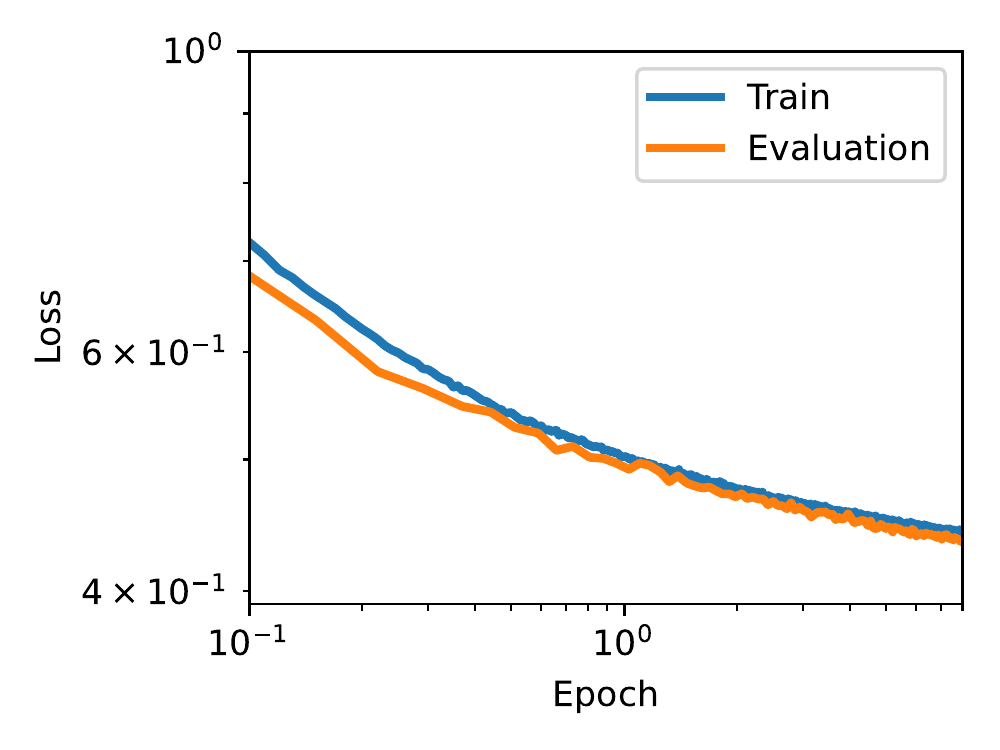}
    \caption{Masked language modeling training and eval loss.}
    \label{fig:mlm_loss}
\end{figure}

AntiBERTy is based on the BERT transformer encoder model \cite{devlin2018bert} implementation from Huggingface \cite{wolf2019huggingface}.  We set the hidden dimension to 512 and the feedforward dimension to 2048.  Our model contains 8 layers, with 8 attention heads per layer.  In total, AntiBERTy contains approximately 26M trainable parameters.  We train the model for 8 epochs over the full dataset, which takes approximately 10 days when parallelized across four NVIDIA A100 GPUs.  Training and evaluation loss are shown in Figure~\ref{fig:mlm_loss}.

\subsection{HIV-1 donor repertoires}
\label{repertoire_info}

For this work, we collected four previously published immune repertoire sequence datasets from donors who developed neutralizing antibodies against the HIV-1 envelope glycoprotein120.  We refer to the four donors using the original identifiers from their respective studies: RU3 \cite{zhou2013multidonor}, IAVI57 \cite{zhou2013multidonor}, IAVI74 \cite{zhou2013multidonor}, and C38 \cite{zhou2015structural}.  For each donor, unsorted B-cells were collected from peripheral blood mononuclear cells (PMBC).  All sequences are of the IGHG isotype.  The number of sequences in each repertoire are given below.

\begin{table}[h]
\centering
\begin{tabular}{ l | c}
\Xhline{2\arrayrulewidth}
\textbf{Donor} & \textbf{Number of sequences}\\
\hline
\texttt{RU3} & 77,067 \\
\texttt{IAVI57} & 67,339 \\
\texttt{IAVI74} & 40,517 \\
\texttt{C38} & 47,670 \\
\Xhline{2\arrayrulewidth}
\end{tabular}
\caption{Size of immune repertoire samples for each donor.}
\label{table:repertoire_sizes}
\end{table}

\subsection{Evo-velocity analysis}
\label{evolocity_scores}

\begin{figure}[h]
    \captionsetup{width=\linewidth}
    \centering
    \includegraphics[width=0.75\textwidth]{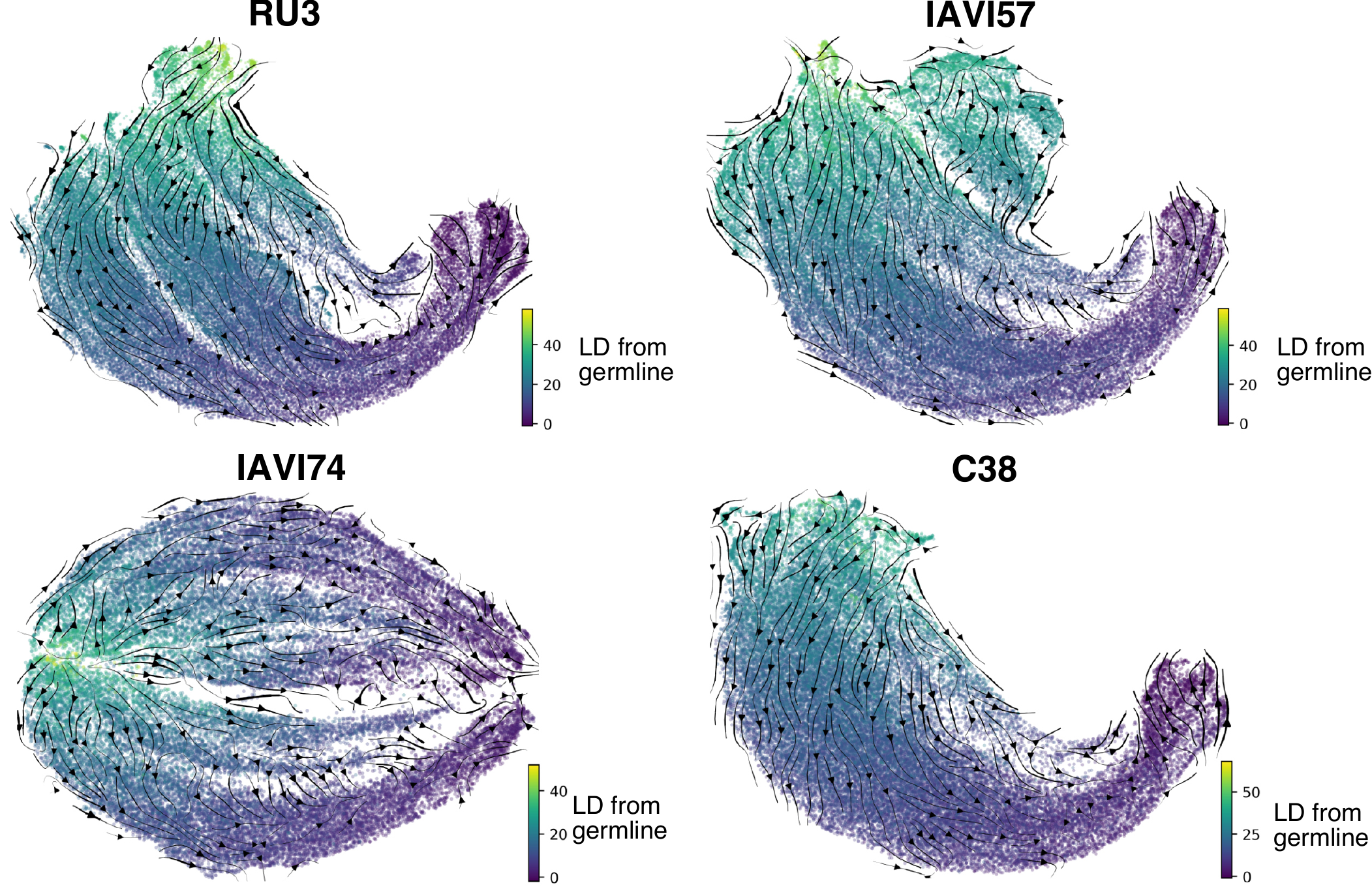}
    \caption{Evo-velocity scores for each repertoire.}
    \label{fig:evolocity_arrows}
\end{figure}

For each repertoire, we computed evo-velocity scores as described by Hie et al. \cite{hie2021evolutionary}.  For each edge in the KNN graph between two sequences $x^a$ and $x^b$, we calculate an evo-velcoity score:

\begin{equation}
    v_{ab} = \frac{1}{M} \sum_{i \in M}{[\log{p(x_i^b \vert \mathbf{z}_i^a)}-\log{p(x_i^a \vert \mathbf{z}_i^b)}]}
\end{equation}

where $M = \{i: x_i^a \neq x_i^b\}$ is the set or residues that differ between $x^a$ and $x^b$, and $z_i$ is the latent representation from AntiBERTy when the $i^{\mathrm{th}}$ residue is masked.  In Figure~\ref{fig:evolocity_arrows}, we plot the evo-velocity scores over the Levenshtein-distance-annotated UMAP embeddings for each repertoire.  We observe that the evo-velocity arrows consistently align towards the germline sequence rather than in the direction of affinity maturation.

\subsection{MIL model parameters}
\label{mil_params}

In Table~\ref{table:mil_params}, we report the parameters for the MIL models used in this work.  In total, each MIL model contains 31.3K trainable parameters.  Models for individual repertoires are trained with batch size 32 and the combined model is trained with batch size 128.  All models are trained using the Adam optimizer \cite{kingma2014adam}.  The learning rate begins at 2e-5 and is decreased according to a cosine annealing schedule.

\begin{table}[h]
\centering
\begin{tabular}{ l | c}
\Xhline{2\arrayrulewidth}
\textbf{Parameter} & \textbf{Value} \\
\hline
$d_\text{emb}$ & 32 \\
$n_\text{heads}$ & 4 \\
$d_\text{head}$ & 16 \\
$d_\text{attn}$ & 32 \\
\Xhline{2\arrayrulewidth}
\end{tabular}
\caption{MIL model parameters}
\label{table:mil_params}
\end{table}

\subsection{MIL model performance metrics}
\label{mil_metrics}

Separate MIL models were trained on sequences from each donor repertoire, as well as a combined dataset of all sequences.  During training, 20\% of sequences were held out for hyperparameter evaluation.  We report performance metrics on this set below.

\begin{table}[h]
\centering
\begin{tabular}{ l | c c c c c}
\Xhline{2\arrayrulewidth}
\textbf{Donor} & \textbf{Accuracy} & \textbf{AUROC} & \textbf{Precision} & \textbf{Recall} & \textbf{F1 Score}\\
\hline
\texttt{RU3} & 0.71 & 0.79 & 0.83 & 0.62 & 0.71 \\
\texttt{IAVI57} & 0.90 & 0.95 & 0.93 & 0.89 & 0.91 \\
\texttt{IAVI74} & 0.90 & 0.96 & 0.95 & 0.88 & 0.91 \\
\texttt{C38} & 0.78 & 0.88 & 0.81 & 0.81 & 0.81 \\
\hline
\texttt{Combined} & 0.75 & 0.84 & 0.74 & 0.81 & 0.77 \\
\Xhline{2\arrayrulewidth}
\end{tabular}
\caption{Performance metrics for MIL models trained on different repertoire datasets.}
\label{table:mil_metrics}
\end{table}

\subsection{Other MIL model attention heads}
\label{mil_heads}

\begin{figure}[h]
    \captionsetup{width=\linewidth}
    \centering
    \includegraphics[width=\textwidth]{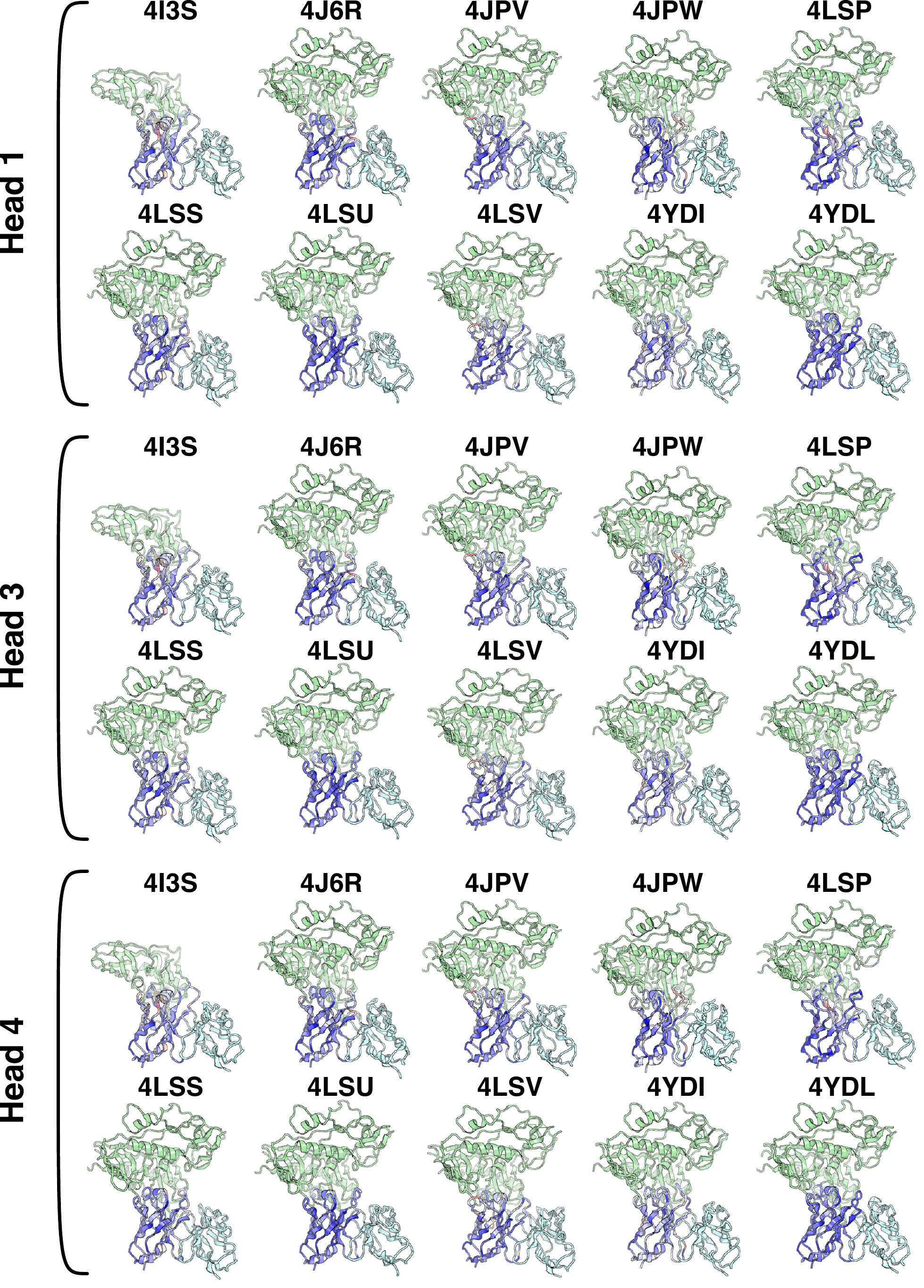}
    \caption{VRC01 antibody complexes annotated with attention from remaining heads.  For each structure, the antibody heavy chain is annotated with residue-level attention from MIL model (increasing from blue to red). The light chain and gp120 antigen are shown in cyan and green, respectively.}
    \label{fig:other_heads}
\end{figure}

The MIL model trained for this work used four attention heads for embedding variable-length sequences to fixed-sized vectors.  Analysis of the second head is provided in the main text (Figure~\ref{fig:figure_3}).  For the remaining heads, we observe attention scattered throughout the antibody (Figure~\ref{fig:other_heads}).  This finding is consistent with observations from previous work that showed substantial framework mutations are necessary to generate functional VRC01 antibodies \cite{klein2013somatic}.

\end{document}